\documentclass[conference]{IEEEtran}
\IEEEoverridecommandlockouts
\usepackage{amsmath,amssymb,amsfonts}
\usepackage{algorithmic}
\usepackage{graphicx}
\usepackage{array, makecell}
\usepackage{textcomp}
\usepackage{float}
\usepackage{xcolor}
\usepackage{color} 
\usepackage{soul} 
\usepackage{hyperref}
\usepackage{biblatex}
\usepackage[normalem]{ulem}

\graphicspath{ {./figures/} }
\bibliography{refs}

\def\BibTeX{{\rm B\kern-.05em{\sc i\kern-.025em b}\kern-.08em
    T\kern-.1667em\lower.7ex\hbox{E}\kern-.125emX}}
\begin{document}

\title{Comparing Bills of Materials\\}

\author{\IEEEauthorblockN{Lucas Tate}
\IEEEauthorblockA{\textit{Pacific Northwest National Laboratory}\\
lucas.tate@pnnl.gov
}
\and
\IEEEauthorblockN{Rebecca Jones}
\IEEEauthorblockA{\textit{Pacific Northwest National Laboratory}\\
}
\and
\IEEEauthorblockN{Doug Dennis}
\IEEEauthorblockA{\textit{Pacific Northwest National Laboratory}\\
}
\and
\IEEEauthorblockN{Tatyana Benko}
\IEEEauthorblockA{\textit{Pacific Northwest National Laboratory}\\
}
\and
\IEEEauthorblockN{Jody Askren}
\IEEEauthorblockA{\textit{Pacific Northwest National Laboratory}\\
}
}

\maketitle

\begin{abstract}
Bills of materials (BOMs) are quickly becoming an effective tool for managing supply chain risk. As more BOMs enter circulation, the ability to compare them will be crucial to understanding how products differ and in managing BOMs from different tools or sources. This paper will describe some of the challenges of comparing BOMs followed by a discussion of several comparison methods.
\end{abstract}


\section{Introduction}

Modern supply chains are increasingly complex. A supply chain for a single product can include multitudes of suppliers, manufacturers, distributors, and more. Common components that drive efficiency and reduce costs also serve to increase the potential damage of any one compromised component.
This complexity poses significant challenges for managing risk because a vulnerability in any one component may have out-sized consequences and the knowledge of what’s inside any given product may be distributed across different companies or across the globe.

Understanding supply chains risks is more important today than it has ever been. Significant vulnerabilities and breaches continue to highlight the growing risks that supply chains face whether they were introduced maliciously or unintentionally. Well known events such as Log4 Shell \cite{kost_log4shell_2023} or Spectre/Meltdown \cite{cybersecurity_and_infrastructure_security_agency_meltdown_2028} demonstrated how weaknesses in components could leave millions of products susceptible to attack. Solar winds \cite{temple-raston_worst_2021} and the recent XZ Utils backdoor \cite{akamai_security_intelligence_group_xz_2024} demonstrated how malicious actors could subvert elements of a supply chain in an attempt to exponentially increase their reach. The increasing persistence and sophistication of supply chain threats requires new tools to combat them.

Recently, BOMs have been gaining traction as a tool to increase our supply chain understanding and help respond to this threat.
A BOM contains details of the components that are used in building a product.
While certainly not a silver bullet, understanding what is inside systems is a first step toward protecting them and responding once protections have failed.
Work around software BOMs (SBOMs) far outpaces other proposed BOMs such as hardware (HBOMs) or artificial intelligence (AIBOMs) with regulations such as Executive Order 14028 \cite{house_executive_2021} and the European Union (EU) Cyber Resilience Act (CRA) \cite{european_commission_proposal_2022} helping to further global adoption. Encouragingly, interest in BOMs has spawned an abundance of new research dedicated to better understanding how to generate, exchange, store, and operationalize BOMs to improve risk management. As BOMs become ubiquitous, we anticipate a growing need for the ability to compare them which will be the focus of this paper.

Due to the nature of BOMs, comparing them is effectively looking at how the composition of two products differs. There are a variety of cases where that might be useful. It may be important to understand how the composition of a product changed with a version update or patch. Another use case might be evaluating a received BOM against an authoritative reference BOM. Comparisons can also be utilized to understand temporal changes that arise in dynamic systems or variation across a family of products. Beyond these use cases, we also find comparison methods useful for identifying inconsistencies in the creation of BOMs themselves. These inconsistencies are discussed later, with \textbf{the irony being that many of these inconsistencies that make comparison difficult are most easily discovered via comparison}.

In this work, we'll start by discussing previous work on comparing BOMs in Section \ref{sec:previous-work}, followed by some of the barriers to comparing BOMs in Section \ref{sec:barriers}.
Section \ref{sec:comparisons} discusses select methods for comparison followed by several examples that illustrate the comparison of two SBOMs in Section \ref{sec:sbom-examples} and two HBOMs in Section \ref{sec:hbom-example}.
Lastly, Section \ref{sec:conclusion} will summarize our conclusions.

\section{Previous Work}\label{sec:previous-work}

As the application of BOMs continues to evolve, particularly in the context of cybersecurity and supply chain management, significant research has been dedicated to understanding and improving how BOMs are generated, compared, and utilized.

Early research on BOMs predominantly focused on enhancing data management techniques to cope with the complexity of large-scale manufacturing environments, such as  implementing a control system to manage BOMs throughout a company \cite{rusk_role_1990} and automating the creation of BOMs using an object-oriented model \cite{chang_manufacturing_1997}.
The object-oriented programming model, similar to the graph method used today, allows for semantic relationships that can create multi-level BOMs with sub-components of components \cite{CHUNG1994321}.
However, since relational databases like SQL are ubiquitous, using them to create BOMs became more common \cite{ NANDAKUMAR198515}.
Algorithms were invented to automate the creation of BOMs by determining which products were in a product order and then pulling the required component information from the database of parts \cite{Aydin_Relational_Database}.
While this method is efficient, flexible and simple, it does not allow for parts explosion or complex computations, especially when analyzing or comparing BOMs \cite{NANDAKUMAR1990471}.

To tackle the growing complexity of intricate products, advancements in BOM structures have been proposed.
The multi-level BOM model is particularly effective in managing software, hardware, or system variations, supporting efficient design and planning \cite{wang_new_2013}. 
Its strengths lies in the ability to handle complex, hierarchical data that can be represented as a graph.

Further development in this area focuses on using graph databases, which store node and relationships, instead of relational databases to integrate product development with production planning \cite{huang_graph_2023}.
BOM management, found at companies such as \href{https://neo4j.com/blog/top-10-use-cases-bill-of-materials/}{Neo4j} and \href{https://www.openbom.com/blog/graphs-networks-and-boms-part-1}{OpenBOM} take this approach of using a graph database.

The main advantage of turning BOMs into graphs is pairing the knowledge of what's inside something with information about how those components relate to each other.
Using graphs, multiple BOMs can easily be combined to gain new information on a larger system, especially when combined with graph visualization tools.
Even when used on a single BOM, graph analysis techniques can provide new insights into a system \cite{cinelli_network_2020}.
Another contribution to the graph-theoretic approach is where BOMs are converted into generic BOM graphs using data mining techniques \cite{Romanowski2002ADM}.
This method leverages graph theory to identify common substructures within BOMs, facilitating the detection of component reuse across different products.
While the ability to uncover hidden patterns and relationships within BOMs is useful, its application could be hindered by computational complexity and graph scalability.

There have been a few attempts to leverage graph theory for BOM comparison.

Graph based similarity analysis has been used to highlight the importance of reducing unnecessary production variations \cite{schmidt_michael_graph-based_2017}
and derived graph similarity metrics have been used to describe the similarity of two BOMs to place them into their product families \cite{romanowski_comparing_2005, kashkoush_product_2016, shih_product_2011}.
Tree reconciliation, matching components in one graph to the components of another, has been used in biology to compare phylogenetic trees and extended to BOMs \cite{kashkoush_matching_2013}.
Similar methods can be applied to BOMs to create new products quickly \cite{ kashkoush_product_2014}.
This is all of the literature we could find on applying graph comparisons to BOMs.

However, graph comparisons have a rich history that could be explored for comparison of BOMs \cite{conte-2004}.
Work has been done on comparing and visualizing trees, a specific type of graph with no cycles \cite{guerra-gomez_treeversity_2013}.
This is might be especially applicable to  HBOMs since they tend to be more hierarchical while SBOMs have a tendency to create loops making it less suitable.
Comparison methods range from similarity metrics, like the ones referenced for BOM graph comparison to unknown node comparisons, which attempt to create a mapping between the nodes of the two graphs \cite{994702}.
In the latter, some algorithms use attributes while others focus solely on the graph structure.
Taking advantage of the attributes is more computationally complex, but is important when comparing BOMs due to the metadata often captured in BOM components.
Since there is not a lot of research on efficient and effective ways of comparing BOMs, graph comparison literature may offer promising methods for future application research.

\section{Barriers to Comparing BOMs}\label{sec:barriers}
BOMs today are extremely heterogeneous which makes subsequent comparison very difficult.
Before undertaking comparison, it’s important to understand some of the sources of variability  \cite{stalnaker_boms_2024, xia_empirical_2023}.
While addressing these differences will be outside of the scope of this paper, considering them will likely be a prerequisite to meaningful comparison.
This list is not exhaustive but captures some of the variability inherent in HBOMs and SBOMs.

\subsection{BOM Standards and Versions}
Currently in the field there are not single authoritative standards describing the structure or contents of an HBOM or an SBOM. For software, the NTIA Minimum elements \cite{the_united_state_departement_of_commerce_minimum_2021} has been an influential guidance document outlining a set of generally accepted minimum elements.
The two leading standards, SPDX \cite{SPDX} and CycloneDX \cite{owasp}, provide detailed schemas that describe a data structure for the capture of SBOM information but the mapping between them can be lossy.
Despite the fact that BOMs have existed in manufacturing for decades, development of HBOM conventions has not reached full maturity. 
In addition to the CycloneDX and SPDX standards, the Information and Communications Technology (ICT) Supply Chain Risk Management (SCRM) Task Force and Department of Homeland Security Cybersecurity \& Infrastructure Security Agency (DHS CISA) released a comprehensive HBOM framework that differs from those standards, although it attempts to provide mappings to them as applicable \cite{cybersecurity_and_infrastructure_security_agency_hardware_2023}.
Even within the same format, major and minor versions describe BOM changes that can impede direct comparison.

\subsection{SBOM Types}
Despite some foundational work defining SBOM types, little has been done to formally differentiate them within real world SBOMs. The result is that two SBOMs for the same software can be markedly different.
As an example, a \textit{source} SBOM created directly from the source code will include named dependencies that are imported or loaded. This will look very different from a \textit{build} SBOM which will describe a specific release and may include information on the build process and produced files.

\subsection{Naming Conventions}\label{sec:naming}
Naming challenges permeate every aspect of BOM generation and despite being a known problem, it is extremely difficult to solve. Software names remain an open challenge. Efforts such as the common platform enumeration (CPE) and package uniform resource locator (PURL) have helped machine-to-machine readability, but they deviate from how people would colloquially refer to software. Other information such as a vendor is complicated by lack of authoritative conventions. As an example `MSFT', 'Microsoft Corporation', and 'Microsoft' are all defensible values but the inconsistent recording makes systematically disambiguating them difficult.
As a last example, hardware component identifiers have a tendency to describe a family of components. This means that sub-strings of the name can still accurately describe components, but  additional characters identify it with increasing specificity. The `AD7579' from Analog Devices describes a LC\textsuperscript{2}MOS 10-Bit Sampling A/D Converter, but `AD7579JN' distinguishes it as having a specific temperature range, integral nonlinearity, and package. Neither name is incorrect, but they utilize different levels of specificity which makes comparison more challenging.

\subsection{Hashing Approaches}
Well known hashing approaches such as MD5, SHA1, SHA256, SHA512 are extremely useful for providing easily matchable fingerprints of files. Their reproducible and static nature make them much more attractive in certain cases than names. One problem is that they are susceptible to dynamic information such as timestamps that often appear in files. Since hashes don't convey why the files are different, it won't be obvious whether the difference is meaningful in a specific comparison. Furthermore, existing standards offer a lot of flexibility in choosing hashing methods which means a different method might have been used from one SBOM to the next reducing comparability.

\subsection{Structure}
Structure in this context describes the relationships between components within a BOM. This structure gives us additional information such as where a dependency is introduced into our software or which board a specific component is mounted on. The problem is that methods for describing these structures are not rigid leading to expected variability in how they are described from one BOM to the next.

\subsection{Scoping}
In this context, scoping describes the boundaries of a BOM; what goes inside a particular BOM and what falls outside. This is a surprisingly hard problem. As an illustrative hardware example, we could describe a Raspberry Pi with an HBOM. If that Raspberry Pi is mounted inside a consumer product, should the HBOM for the consumer product include an external reference to the Raspberry Pi HBOM? Should it duplicate the information from the Raspberry Pi? From a software perspective if a software application requires the use of a shared library in the operating system, should that be included in the SBOM? The lack of a clear and accepted answer to these scoping questions result in variability that needs to be considered.

\subsection{Quantities}
Quantities are an interesting property that appear in HBOMs.
Rather than listing a component \textit{n} times, we can indicate how many of them are present with an integer value.
However, if one HBOM opts to list the components individually and another HBOM leverages the quantity field, then you have to rectify these different representations when conducting a comparison.

\subsection{Order}
BOMs are unordered. This makes sense because there isn't a correct order to describe components. This property however immediately adds a lot of variability to the files which poses some challenges for simple comparative methods such as tabular comparison especially in conjunction with name variation that will stymie attempts to sort the data.

\section{Comparing BOMs}\label{sec:comparisons}

Unfortunately there is no single method that can be used to compare BOMs. Instead, strategies need to be specifically chosen based on the data available in the BOMs in conjunction with consideration of the questions that need to be addressed.
In a simple example, if we want to understand the difference in \textit{licenses} between two BOMs that utilize a well-formed ontology such as the SPDX License List \cite{SPDX}, a set comparison using exact match of the values can be employed successfully.

In other cases, understanding quantities can be important. If we consider two HBOMs and want to understand how the components differ, we may opt for a list comparison of \textit{component names} which will tell us if there are different components, as well as whether there were a different number of them used.
We know from earlier discussion that \textit{component names} can have a lot of inherent variability, so depending on the consistency of the data, a fuzzy matching technique may be needed.
Fuzzy matching allows for some threshold of leniency in matching values that are `close enough' at the expense of possibly making errant matches.

Direct element comparisons are not the only useful comparisons to be made. Creative use of redundant or complimentary information can be exploited to great effect. This is especially useful when comparing SBOMs where \textit{component names}, \textit{hashes}, \textit{cpes}, or \textit{purls} can be used together to gain additional insights. As an example, matching \textit{hashes} provide some level of guarantee that the contents of a software component match. When compared to \textit{component names} this can identify interesting situations where 1) \textit{component names} are the same, but the contents differ, 2) \textit{component names} are different, but the contents are the same, or 3) offer consensus between \textit{component names} and \textit{hashes}. These comparisons can often uncover unexpected results that are invaluable for assessing quality and consistency of BOMs.

So far, the comparisons that have been discussed implicitly assume a comparison of two similar BOMs, but other comparisons can be useful as well. 
SBOM practitioners will likely be intimately familiar with the variability of generation tools.
Despite the monumental efforts around standardization, SBOMs tend to vary greatly from one tool to the next.
This can happen for many reasons, but some examples include inconsistent assumptions, different methods or levels of technical ability, different opinions on the boundaries of an SBOM, and opinions on whether transitive dependencies should be included.
Further exacerbating the issue is the fact that the details driving the variability are often proprietary or black-box.
By comparing the lists and sets of elements within the BOMs it is possible to gain insights into design choices and accuracy of various tools.

BOMs of different size can also be compared. This can suggest that the components of one BOM are a subset or contained within another BOM.
It is also possible to explore subsets of a BOM; in a system with built-in redundancy it may be useful to look at how duplicated modules or sub-assemblies compare.
Much of the previous work done on comparing BOMs has been used to cluster or identify product families that contain similar components in a similar structure. 
Comparing products within a family can lead to quicker generation of new products, as well as streamlining supply chain processes.

\subsection{List and Set Comparisons}
A straightforward approach to comparing two BOMs is by simply comparing lists. Due to the popularity of JSON and XML file types for use in BOMs, this will often require parsing and/or flattening of the data to obtain the unordered lists. An example this could be comparing all the \textit{component names} in one BOM to the \textit{component names} in another BOM. List comparisons aid in understanding differences and quantities of components which can be particularly useful if multiples of a single component are present. It should be noted that there are cases where comparing multiple elements simultaneously is necessary. For example, two manufacturers could use the same name for a particular component in which case it may be more useful to compare the manufacturer and name at the same time so as to differentiate one component from the other.

Beyond simple lists it can also be useful to only consider the unique values or sets. This representation sacrifices information about the frequency of values, but can greatly reduce the burden of comparison. Set comparison might be useful when looking at something like \textit{licenses} where knowing that a license appears in \textit{n} dependencies is probably less important than just having the list of unique licenses.

\subsection{Graph Comparisons}
Graph matching techniques can provide useful insight into the comparison of two BOMs.
The list comparison approach does not take advantage of the relationships which are present between elements in a BOM.
For example, if there are duplicates of a chip on a piece of hardware, set comparisons will not capture that information, while a graph comparison will.
Importantly, it will also show where the chips are physically in the hardware.
This can be done through text like the list comparison or crucially, visualization techniques that are intuitively easy to understand.

In order to take advantage of this visualization ability, the BOMs are converted to a graph by making the components into nodes and the relationships between two elements as edges.
Information about each element can be recorded in the graph by associating node attributes, and relationship types can be given by edge attributes. 
Then a node mapping where one set of nodes is mapped to the other is created by using the edges and the node attributes.

\section{SBOM Example}\label{sec:sbom-examples}

To illustrate the SBOM comparison methods, we used Trivy to generate SBOMs from two versions (3.6.4 and 3.7.0) of  \href{https://github.com/thingsboard/thingsboard}{Thingsboard}, an open-source IoT platform for data collection, processing, visualization, and device management. Thingsboard is largely written in Java and uses Maven to manage the project. Several modifications were made to the SBOMs. First, duplicate software components that shared the same purl and metadata were collapsed to a single component. If relationships existed to the duplicates that were removed, they were tied to the remaining copy. While the exact nature of the duplication in these SBOMs wasn't clear, it should be noted that removing them could hinder certain analyses such as finding multiple copies of a dependency. Next we opted to remove all the npm front-end dependencies. This was only done to make the SBOMs a little smaller for illustration.

The comparison of the resulting SBOMs (results in Table \ref{tab:sbom-set}) showed that out of the $230$ components, there were $214$ unique names detected in version 3.6.4 and $218$ unique names from the $234$ components in version 3.7.0.
Interestingly, the number of unique purls was also four apart: $170$ to $174$.
We note that the difference between the number of components and purls was due to  $60$ components that did not contain purls in the SBOM.
The number of unique purls is identical to the total purls which is expected after the deduplication, meaning that each purl appears only once.
There were $10$ component names that were duplicated; each one had a different unique purl, with a total of $16$ duplicates.
Looking at the comparisons of the unique names, there were $201$ names that appeared in both SBOMs.
Finally, doing a Jaro-Winkler string comparison with a threshold of greater than $0.85$ on the node names between the two sets resulted in a total of $887$ matches.

\begin{table}[h]
\begin{center}
\caption{Comparison Results for SBOM where 3.6.4 (Only) indicates the difference between 3.6.4 and 3.7.0}
\begin{tabular}{|c| c| c| c| c|}
\hline
& 3.6.4 & 3.7.0 & 3.6.4 (Only) & 3.7.0 (Only)\\
\hline
Name & 230 & 234 & 13& 17\\ 
Unique Names & 214 & 218 & 13& 17 \\
 Purls & 170 & 174 & 158&162\\ 
 Unique Purls &170 &174 &158 &162\\
\hline
\end{tabular}
\label{tab:sbom-set}
\end{center}
\end{table}

Comparing two SBOMs does not have to be constrained to differences in components.
We also considered whether the licenses reported were different between the generated SBOMs. Because the primary interest is whether there are any different licenses to consider, using set comparison (comparing the unique values) of recorded license would seem the obvious choice. Version 3.7.0 contains only a single unique license: Apache-2.0.
However, version 3.6.4 contains two unique licenses: Apache-2.0 and MIT. While this may look like version 3.6.4 is more complete, it is also important to know that only six dependencies in the version 3.6.4 SBOM had a license recovered by the tool. Version 3.7.0 had four dependencies with recovered licenses.
This largely indicates that neither SBOM has a complete picture of the licensing exposure in Thingsboard.

The potential lack of information prompted us to manually review the licenses for the listed dependencies. 
We discovered that both versions of the software contained several other licenses such as the Eclipse Public License (EPL) and the Lesser GNU Public License (LGPL). These licenses have additional disclosure and representation requirements that may not be satisfied with the same rules as Apache-2.0 or MIT. Additionally, it is not inconceivable that an organization may apply additional scrutiny to licenses from the GNU Public License (GPL) family and would want to know that LGPL code is being used.
While set comparison identified a notable difference in recorded licenses, a list comparison would have highlighted how few of the components captured license information.

We also examined the organizations of the dependencies.
They were detected by using the first two segments of the package name in the purls.
For example, given the purl ``pkg:maven/com.example.foo@1.2.3", the organization is ``com.example".
There were fifty unique organizations in version 3.6.4 and 52 in version 3.7.0.
Excluding Java standard library packages, we found that when moving to version 3.7.0 Thingsboard gained four additional external organizations and lost one. This information could be useful for situational awareness or subsequent corporate analysis where some producers may imply an increased/decreased level of assurance.


For the graph comparison, we converted the SBOMs into graphs using components as nodes and dependencies as edges. We  then merged them on the name field using a depth-first search matching algorithm \cite{10418603} with exact matching.
A quick calculation shows us that $217$ nodes were matched, $13$ appeared only in version 3.6.4 and $17$ appeared in version 3.7.0, matching the comparison in Table \ref{tab:sbom-set}.
The visualization of the compared graphs did not add to the analysis and was therefore not included.
To understand if the identified differences between the two SBOMs  was due to small variance in the names, we employed fuzzy matching. It should be noted that with a priori knowledge of the name structure, some fuzzy matching approaches may be more successful, but here we naively employed Jaro-Winkler on the node names with an arbitrary threshold of $.85$ and found that there are $7$ similar packages (see Table \ref{tab:sbom}). While we note that the graph wasn't particularly useful for visualization, the added constraint of the structure reduced the number of possible matches by $880$ because rather than just finding similar names, the structure requires the names to also be in the same place as defined by the relationships in the SBOM.
\begin{table}[h!]
\begin{center}
\caption{Similar names of packages in each Thingsboard version with the differences highlighted.}
\begin{tabular}{| c| c| }
\hline
3.6.4 & 3.7.0\\
\hline
bcpkix-jdk\hl{15}on & bcpkix-jdk\hl{18}on  \\ 
 bcprov-jdk\hl{15}on & bcprov-jdk\hl{18}on \\  
commons-collections & commons-collections\hl{4}\\
hypersistence-utils-hibernate-\hl{55} &hypersistence-utils-hibernate-\hl{63}\\
\hl{javax}.annotation-api & \hl{jakarta}.annotation-api\\
swagger-annotations &swagger-annotations\hl{-jakarta}\\
spring\hl{fox}-boot-starter & spring-boot-starter\hl{-webflux}\\
\hline
\end{tabular}
\label{tab:sbom}
\end{center}
\end{table}

In comparing these SBOMs, a combination of methods proved useful. List and set comparisons provided useful characterization of the SBOMs and identified some license irregularities. The graph method allowed us to ignore dependencies with similar names and focus on the differences we are more interested in, which were updated dependencies.


\section{HBOM Example}\label{sec:hbom-example}
\begin{figure}[htbp]
\centerline{\includegraphics[width=\linewidth]{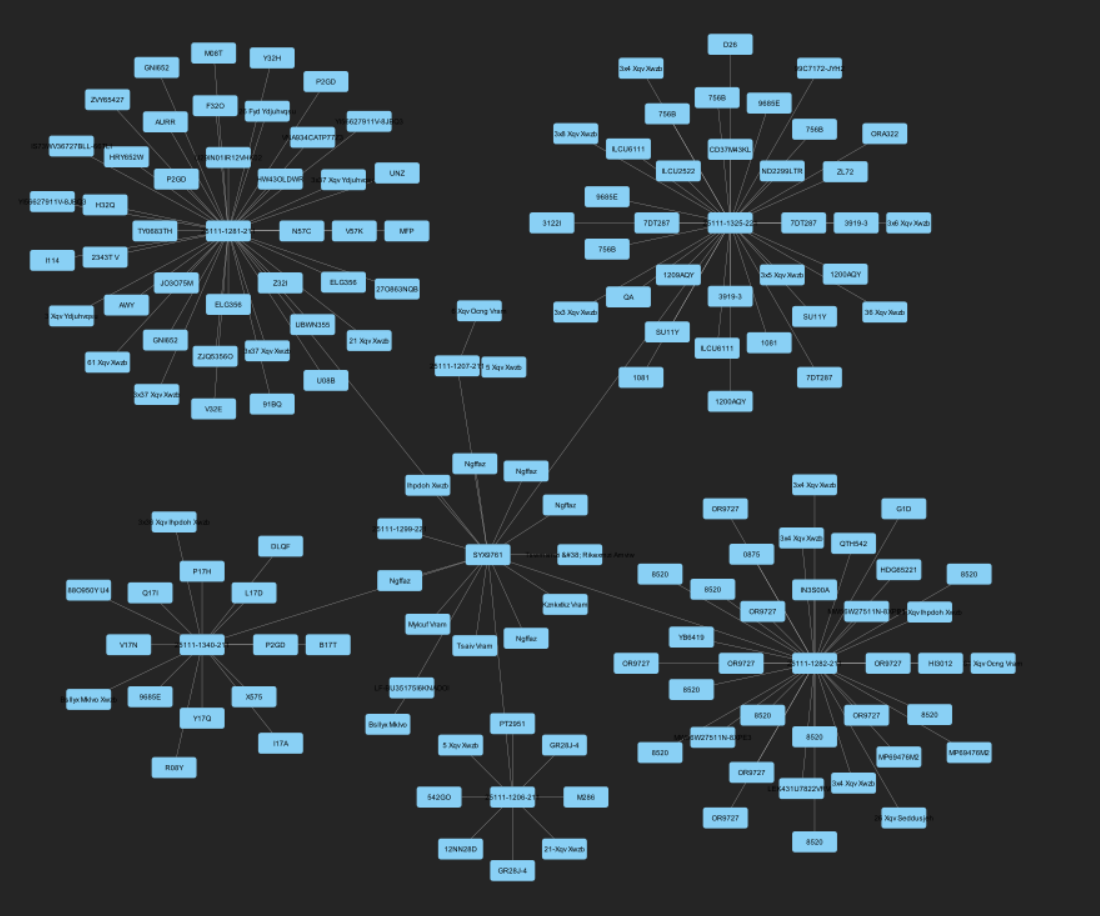}}
\caption{Visualization of HBOM 1.}
\label{fig:hw1}
\end{figure}

\begin{figure}[htbp]
\centerline{\includegraphics[width=\linewidth]{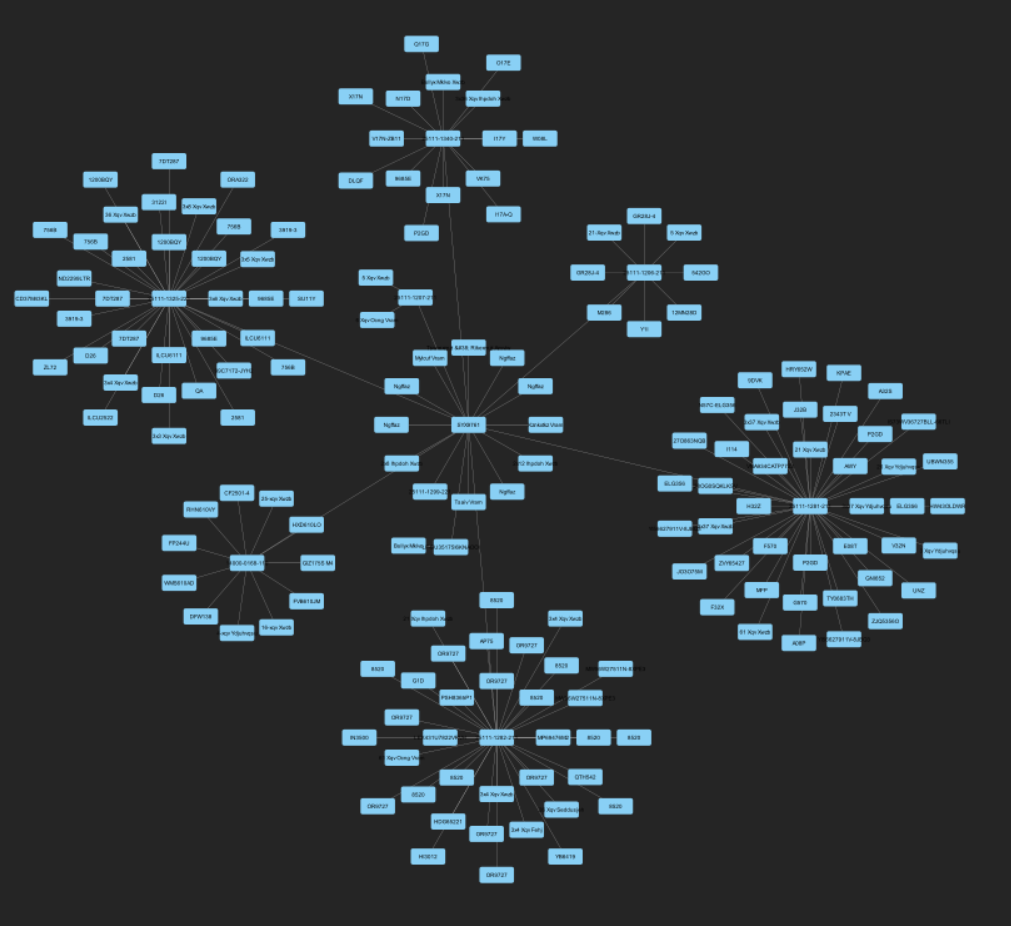}}
\caption{Visualization of HBOM 2}
\label{fig:hw2}
\end{figure}

In this example, two HBOMs were created from two distinct instances of the same hardware product.
The identities of the devices and components have been obfuscated, but the real characteristics of the comparison were preserved.
Visualizations of the two graphs are shown in Figures \ref{fig:hw1} and \ref{fig:hw2}.

A list comparison of the component names for the HBOMs immediately conveys differences shown in Table \ref{tab:hbom-set}.
Despite being the same product, we see 35 components that only appear in HBOM 1 and 46 components that only appear in HBOM 2.
Because the comparison of unique names reflects different counts, we can infer that some of the differences include components that appeared multiple times.
In reviewing the differences, the most noteworthy finding was that one of the components was a circuit board which was especially surprising. Fuzzy matching in this case was not particularly useful because it flagged 743 possible matches which is difficult to sift through.

A comparison of the vendor revealed that there were 17 unique vendors in HBOM 1 and 14 unique vendors in HBOM 2.
with  only 13 vendors shared between the two. We offer no explanation as to why the vendors differ, but interestingly despite the BOMs representing the same product, the component supply chain looks different and may result in varying levels of risk exposure.

\begin{table}[h]
\begin{center}
\caption{Comparison Results for HBOM Example Where HBOM 1 (only) indicates the difference between HBOM 1 and HBOM 2}
\begin{tabular}{|c| c| c| c| c|}
\hline
& HBOM 1 & HBOM 2 & \makecell{HBOM 1\\(Only)} & \makecell{HBOM 2\\ (Only)} \\
\hline
Name & 156 & 169 & 35 & 46\\ 
Unique Names & 99 & 108 & 17 & 26\\

\hline
\end{tabular}
\label{tab:hbom-set}
\end{center}
\end{table}

\begin{figure}[htbp]
\centerline{\includegraphics[width=\linewidth]{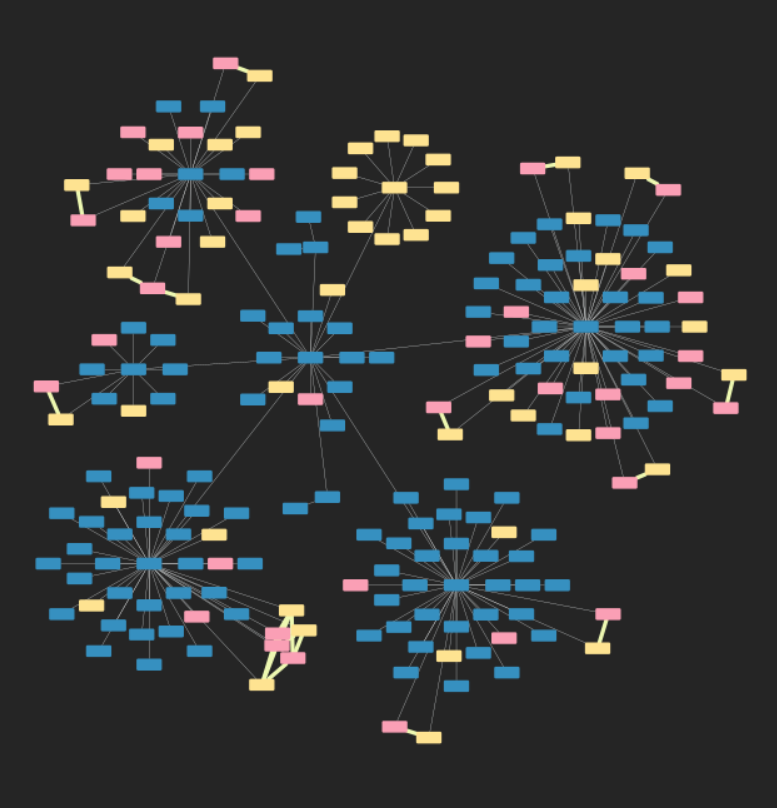}}
\caption{Visualization of Merged Graph. Blue represents nodes found in both HBOM 1 and HBOM 2. Pink nodes are only found in HBOM 1 and yellow nodes are only in HBOM 2. Yellow edges indicate nodes where the names fuzzy matched.}
\label{fig:merged-hbom}
\end{figure}

Repeating the same graph comparison approach as in Section \ref{sec:sbom-examples}, we generated the merged graph shown in Figure \ref{fig:merged-hbom}.
Blue nodes indicate that the node names matched exactly, while the thick yellow edges indicate the node names matched approximately (Jaro-Winkler with a threshold of $.85$).

This visualization is immediately useful. The most notable difference is the presence of a yellow circle in the top of Figure \ref{fig:merged-hbom}.
This turned out to be the additional circuit board that was unexpectedly present in one of the devices, and we can quickly understand which of the different components correlate to the addition of that board.
The remaining differences are drastically reduced by enforcing the graph structure (e.g. the component must be on the same board). The differences identified fell into one of three categories:
\begin{enumerate}
\item The component name was transcribed incorrectly. For example, the names were recorded as $IN3S00A$ and $IN3500$, where a $5$ is switched for an $S$.
\item The difference was real and describes a component that had been switched out in production with a different but equivalent component.
\item The name in one HBOM was recorded with more specificity than the name of the equivalent component in the other HBOM. One component was named $V17N$ and one was recorded as $V17N-ZB11$.
\end{enumerate}

Fuzzy match on the component names using the graph was particularly useful for this comparison. Instead of 743 possible matches, there are only 21 matches, identifying transcription errors and incremental component variations with high precision.
As with the SBOM comparison, leveraging multiple comparison methods proved to be useful, but notably the graph provided significantly increased utility in the hardware example. 


\section{Conclusion}\label{sec:conclusion}
This paper provides an introductory discussion of the methods and challenges associated with comparing BOMs. Despite the recent abundance of energy and research in BOMs by government, industry, and academia, tools and methods to effectively compare BOMs lag behind. There is no single method of comparison that can effectively compare BOMs today. List, set, and graphical comparisons are complimentary and contribute to a foundational capability. As reference BOMs become more readily available, comparison methods will be vital to leveraging them. Ultimately BOM adoption, spurred by policy and regulation will continue to grow. The ability to compare BOMs will be essential for understanding and reasoning about BOMs for supply chain risk management. 
\section*{Acknowledgment}

The authors wish to thank the Department of Energy (DOE) Cybersecurity, Energy Security, and Emergency Response (CESER) and the Cyber Testing and Resilience of Industrial Control Systems (CyTRICS) Program including Idaho National Laboratory (INL), Lawrence Livermore National Laboratory (LLNL), National Renewable Energy Laboratory (NREL), Oakridge National Laboratory (ORNL), and Sandia National Laboratory (SNL). Information Release: PNNL-SA-202952.

\newpage
\printbibliography
\end{document}